\documentclass{article}

\usepackage[final,nonatbib]{nips_2018}

\usepackage[utf8]{inputenc} % allow utf-8 input
\usepackage[T1]{fontenc}    % use 8-bit T1 fonts
\usepackage{url}            % simple URL typesetting
\usepackage{booktabs}       % professional-quality tables
\usepackage{amsfonts}       % blackboard math symbols
\usepackage{nicefrac}       % compact symbols for 1/2, etc.
\usepackage{microtype}      % microtypography
\usepackage{textcomp}

\usepackage{mathtools}      % mathtools loads amsmath
\usepackage{amssymb,amsthm}
\usepackage{stmaryrd}
\usepackage[mathcal]{euscript}
\usepackage{bm}
\usepackage{url}
\usepackage{xcolor}

\usepackage{hyperref}       % hyperlinks
\hypersetup{
  linktoc = page,
  pdfpagemode = UseNone,
  colorlinks,
  linkcolor={red!50!black},
  citecolor={blue!50!black},
  urlcolor={blue!80!black}}
\usepackage[capitalize]{cleveref}

\title{How to Profile Privacy-Conscious Users in Recommender Systems}

\author{%
  Fabrice Benhamouda\\
  IBM Research\\
  Yorktown Heights, NY, USA
  \And
  Marc Joye\\
  OneSpan\\
  Brussels, Belgium}

\newcommand*{\ie}{i.e.}
\newcommand*{\eg}{e.g.}

\newcommand*{\TSet}{\mathcal{D}}
\newcommand*{\SSet}{\mathcal{S}}
\newcommand*{\bbbr}{\mathbb{R}}
\newcommand*{\bbbz}{\mathbb{Z}}
\newcommand*{\Z}{\bbbz}

\newcommand*{\Zn}{\bbbz_n}
\newcommand*{\GL}{\mathop{\mathsf{GL}}}
\renewcommand*{\vec}[2][0mu]{\bm{#2\mkern-#1}\mkern#1}
\newcommand*{\smallset}[1]{{\{#1\}}}

\DeclareMathOperator*{\argmin}{arg\,min}

\DeclareMathOperator{\lcm}{lcm}
\newcommand*{\KeyGen}{\mathop{\mathsf{KeyGen}}}
\newcommand*{\Enc}{\mathop{\mathsf{Enc}}}
\newcommand*{\Dec}{\mathop{\mathsf{Dec}}}
\newcommand*{\msgset}{\mathcal{M}}
\newcommand*{\pk}{\mathsf{pk}}
\newcommand*{\sk}{\mathsf{sk}}
\newcommand*{\secpar}{\kappa} 
\DeclarePairedDelimiter{\dbracket}{\llbracket}{\rrbracket}
\newcommand*{\HEnc}[2][\relax]{%
  \ifx#1\relax\dbracket{#2}\else\dbracket{#2}_{#1}\fi}
\makeatletter
\renewcommand*{\boxplus}{\mathbin{\ooalign{%
      $\m@th\boxempty$\cr
      \hidewidth$\m@th\hbox{+}$\hidewidth\cr      
    }}}
\makeatother
\newtheorem{theorem}{Theorem}%[section]
\Crefname{theorem}{Theorem}{Theorems}
\theoremstyle{remark}

\newtheorem{example}[theorem]{Example}

\makeatletter
\let\orig@xsect=\@xsect
\def\@addperiod{.}
\renewcommand{\@xsect}[1]{%
  \@tempskipa #1\relax 
  \ifdim \@tempskipa >\z@ 
  \else 
    \global\@noskipsectrue 
    \g@addto@macro{\@svsechd}{\@addperiod}
  \fi
  \orig@xsect{#1}}
\makeatother

\begin{document}

\maketitle

\begin{abstract}
  Matrix factorization is a popular method to build a recommender
  system.  In such a system, existing users and items are associated
  to a low-dimension vector called a \emph{profile}.  The profiles of
  a user and of an item can be combined (via inner product) to predict
  the rating that the user would get on the item.  One important issue
  of such a system is the so-called cold-start problem: how to allow a
  user to learn her profile, so that she can then get accurate
  recommendations?

  While a profile can be computed if the user is willing to rate
  well-chosen items and/or provide supplemental attributes or
  demographics (such as gender), revealing this additional information
  is known to allow the analyst of the recommender system to infer
  many more personal sensitive information.  We design a protocol to
  allow privacy-conscious users to benefit from
  matrix-factorization-based recommender systems while preserving
  their privacy.  More precisely, our protocol enables a user to learn
  her profile, and from that to predict ratings without the user
  revealing any personal information.  The protocol is secure in the
  standard model against semi-honest adversaries.
\end{abstract}

%%%%%%%%%%%%%%%%%%%%%%%%%%%%%%%%%%%%%%%%%%%%%%%%%%%%%%%%%%%%%%%%%%%%%%%%%%%%%%%
\section{Introduction}
%%%%%%%%%%%%%%%%%%%%%%%%%%%%%%%%%%%%%%%%%%%%%%%%%%%%%%%%%%%%%%%%%%%%%%%%%%%%%%%

Matrix factorization~\cite{CanRec09,KesMonOh08} is a popular method to
build a recommender system.  As exemplified by the \textsl{Netflix
  Prize} competition~\cite{NetflixPrize}, it has become a dominant
technology within collaborative-filtering recommenders.  Matrix
factorization provides a better predictive accuracy compared to
classical neighborhood methods while at the same time is scalable and
offers much flexibility for modeling a variety of real-life
situations~\cite{KorBelVol09}.

\paragraph{The cold-start problem}
A major problem facing collaborative-filtering recommender systems is
how to provide recommendations when rating data is too sparse for a
subset of users or items.  As a special case, the so-called
\emph{cold-start problem}~\cite{SPUP02} is how to make recommendations
to new users who have not yet rated any item or to deal with new items
that have not yet been rated by users.

The cold-start problem is usually addressed by incorporating
additional input sources to compensate for the lack of rating data.
In addition to ratings, the analyst may for example collect certain
user attributes, such as gender, age, or other demographic
information~\cite{AdoTuz05,Koren08}.

Another approach for dealing with the cold-start problem is to ask
users to rate a minimum number of (well chosen)
items~\cite{RACLMKR02}.  

% Specifically, when a new user, say user $i$,
% wishes to use the service, she submits a batch of $s$ ratings
% \[
% r_{i,j_1},\thickspace r_{i,j_2},\thickspace \dotsc,\thickspace r_{i,j_s}
% \]
% for a subset of $s$ items $j_1, j_2, \dotsc, j_s \in \{1, \dotsc,
% M\}$, with $s \ge d$.  Upon receiving these ratings, the analyst can
% estimate the profile $\vec{u_i}$ of user $i$ through
% \begin{equation}\label{eq:estimate_ui}
%   \min_{\vec{u_i}\in\bbbr^d} \sum_{j \in\{j_1, \dotsc,
%     j_s\}}\bigl(r_{i,j} - \langle \vec{u_i}, \vec{v_j}
%   \rangle\bigr)^2\thinspace, 
%   % + \nu \, {\|\vec{u_i}\|}^2\thinspace, 
% \end{equation}
% % for some constant $\nu \ge 0$.
% and subsequently predict ratings for items $j \notin \{j_1, \dotsc,
% j_s\}$, using Eq.~\eqref{eq:prediction}.

\paragraph{Inference attacks}
Relying on additional input sources to address the cold-start problem
may be difficult to deploy in practice as privacy-conscious users may
be reluctant to supply some of their attributes.  The second approach
does not require to collect extra information beyond the ratings and
is very efficient.  Unfortunately, the
% ratings received for solving Eq.~\eqref{eq:estimate_ui} may reveal a
% lot of information about user $i$
additional received ratings may reveal a lot of information about a
user to the analyst. Recent research has indeed demonstrated that this
can be used by the analyst to infer private user attributes such as
political affiliation~\cite{KosStiGra13,SAPBDKOT13}, sexual
orientation~\cite{KosStiGra13}, age~\cite{WBIT12},
gender~\cite{SAPBDKOT13,WBIT12}, and even drug use~\cite{KosStiGra13}.
Further privacy threats are reported in
\cite{BWIT14,CKNFS11,SP:NarShm08}.

A natural question therefore raised in \cite{IMWBFT14} is how a
privacy-conscious user can benefit from recommender systems while
preventing the inference of her private information.

\paragraph{Contributions}
In this paper, we show how a privacy-conscious user can learn her
profile without revealing any information to the analyst.  The
protocol is practical and proven secure against semi-honest
adversaries.  The communication complexity of the protocol only grows
with the square-root of the number of items.

Once the privacy-conscious user learns her profile $\vec{u}$, she can
run a straightforward protocol to learn the predicted rating of any
item $j$ in the database.  This indeed only requires to compute the
inner product between the user profile $\vec{u}$ (known to the user)
and the item profile $\vec{v_j}$ (known to the
analyst)~\cite[Sect.~4.1]{CCS:NIWJTB13}.

\paragraph{Related work}
In~\cite{IMWBFT14}, Ioannidis et al.\ propose a learning protocol
which enables the user to prevent the analyst from learning some
(previously defined) private user attributes.  This protocol perfectly
hides these chosen attributes to the analyst, in an
information-theoretic way.  The authors also prove that no such
protocol can be more accurate, when the analyst ends up knowing the
resulting profile, nor can disclose less information for the same
accuracy.

Unfortunately, this protocol has also several drawbacks, most of them
inherent to the fact it is information-theoretically secure and does
not rely on computational assumptions.  First, this protocol still
needs to disclose some information about the analyst database to
everybody.  Second, this protocol is not as accurate as a
non-privacy-preserving protocol would be.  This is inherent to the
fact that Ioannidis et al.\ restricted themselves to protocols where
the analyst learns an approximate profile of the user at the end, so
that the resulting user profile shall not contain any information
about the private attribute.  Third, it can only hide a small fixed
set of attributes: all attributes which are not explicitly hidden may
be recovered by the analyst.  And it may be hard for a user to decide
which attributes are really important to her, due to the wide range of
possible attributes.  Finally, the analyst needs to ask
users\footnote{In the simplest scenario, we have to restrict to
  non-privacy-conscious users.  But it would also be possible to
  compute item profiles using privacy-preserving matrix
  factorization~\cite{CCS:NIWJTB13}.} to reveal which attributes they
deem private.  This may not only bother a lot these users, but also
brings up the question of the reliability of these data.  No user will
be likely admitting she is a drug addict, for example, even if she is
ensured that this data will not be disclosed.\footnote{Notice in
  particular that, in the privacy-preserving matrix factorization
  protocol in~\cite{CCS:NIWJTB13}, in case of collusion between the
  CSP (Crypto Service Provider) and the analyst, it is possible to
  recover all data sent by the user.  This means that governmental
  agencies may force the recommendation systems to disclose these
  private user attributes.}

%%%%%%%%%%%%%%%%%%%%%%%%%%%%%%%%%%%%%%%%%%%%%%%%%%%%%%%%%%%%%%%%%%%%%%%%%%%%%%%
\section{Preliminaries}
\label{sec:preliminaries}
%%%%%%%%%%%%%%%%%%%%%%%%%%%%%%%%%%%%%%%%%%%%%%%%%%%%%%%%%%%%%%%%%%%%%%%%%%%%%%%
$\bbbr$ is the field of real numbers. For any integer $n$, $\bbbz_n$
is the ring of integers modulo $n$, while $\bbbz_n^*$ is its
multiplicative group. Vectors are always column vectors and are
denoted as $\vec{u}$ or $\vec{v_j}$.  Matrices are denoted with
capital letters.
% A positive function $f$ is negligible, if for any positive integer
% $k$, for any large enough positive integer $\secpar$:
% $f(\secpar) \le 1/\secpar^k$.

\subsection{Cryptographic tools}
%~~~~~~~~~~~~~~~~~~~~~~~~~~~~~~~
\paragraph{Public-key encryption}
A \emph{public-key encryption scheme} is defined by three algorithms:
$\KeyGen$, $\Enc$, and $\Dec$.  $(\pk,\sk) \gets \KeyGen(1^\secpar)$
generates a matching pair of public key $\pk$ and secret key $\sk$,
given a security parameter $1^\secpar$ (unary notation).  The public
key $\pk$ is used to encrypt a message $x \in \msgset$ into a
ciphertext $c$:  $c \gets \Enc(\pk,x)$.  The secret key $\sk$ is used
to decrypt a ciphertext $c$:  $x \gets \Dec(\sk,c)$.  We assume that
the encryption scheme is perfectly correct and semantically secure
(\ie, IND-CPA)~\cite{GolMic84}.

% An encryption key needs to satisfy two properties:
% \begin{itemize}
% \item Perfect correctness: for any
%   $(\pk,\sk) \gets \KeyGen(1^\secpar)$, for any $x \in \msgset$, for
%   any $c \gets \Enc(\pk,x)$, $\Dec(sk,c) = x$.
% \item Semantic security (aka, IND-CPA): for any polynomial-time
% adversary $\advA$, there exists a negligible function $\negl$ such
% that: 
%   \[ \Pr\left[\begin{array}{l}
%                 (\pk,\sk) \gets \KeyGen(1^\secpar);\; b \gets \{0,1\};\;\\
%                 (\state,x_0,x_1) \gets \advA(1^\secpar,\pk);\;
%                 \advA(\state, \Enc(\pk,x_b)) = b 
%               \end{array}
%             \right] \le \negl(\secpar)\enspace. \]
% \end{itemize}

\paragraph{Homomorphic encryption}
An \emph{additively homomorphic encryption scheme} is such that the
message set $\msgset$ is an additive group, and there exists a
randomized operation $\boxplus$ such that
$\Enc(\pk,x) \boxplus \Enc(\pk,y)$ is distributed identically to a
fresh ciphertext of $x + y$.  This operation can be extended to a
scalar multiplication by an integer $k$: $k \boxdot \Enc(\pk,x)$ is a
fresh ciphertext of $k \cdot x$; that is, $x + x + \cdots + x$ ($k$
times).

To simplify the notation, we will sometimes use $\HEnc[\pk]{x}$ for
$\Enc(\pk,x)$ and omit $\pk$ when clear from the context. We so have
$\HEnc{x+y} = \HEnc{x} \boxplus \HEnc{y}$ and
$\HEnc{k\cdot x} = k \boxdot \HEnc{x}$.\par\smallskip

% % \begin{example}[Okamoto-Uchiyama encryption scheme]
%   We recall the Okamoto-Uchiyama encryption scheme\cite{EC:OkaUch98a},
%   which is an homomorphic encryption scheme that is semantically
%   secure under the $p$-subgroup assumption.
%   $(\pk,\sk) \gets \KeyGen(1^\secpar)$ generates two large primes $p$
%   and $q$, set $n = p^2 q$, choose an element $g \in \Z_n^*$
%   uniformly at random such that $g^{p-1} \neq 1 \bmod p^2$, and set
%   $h = g^n \bmod n$. The public key is $\pk = (n,g,h)$, while the
%   secret key is $\sk = (p, q, g)$. $c \gets \Enc(\pk, x)$ picks a
%   uniformly random $r \gets \Z_n$ and return $c = g^x j^r \bmod n$.
%   $x \gets \Dec(\sk, c)$, return
%   $x = L(c^{p-1}\bmod p^2) / L(g^{p-1} \bmod p^2) \bmod p$ where
%   $L(a) = (a-1) / p$.

%   The scheme is homomorphic:
%   $c_x \boxplus c_y = c_x \cdot c_y \cdot h^r \bmod n$, where
%   $r \gets \Z_n$.  The group of messages is $\Zp$. We remark however
%   that $p$ is not known to the encryptor.
% \end{example}

\begin{example}[Paillier encryption scheme]\label{ex:paillier}
  We recall the Paillier encryption scheme~\cite{EC:Paillier99}, which
  is an homomorphic encryption scheme that is semantically secure
  under the Decisional Composite Residuosity (DCR) assumption.
  $(\pk,\sk) \gets \KeyGen(1^\secpar)$ generates two large
  equal-length primes $p$ and $q$, computes $n=pq$, and sets
  $\lambda = \lcm(p-1,\, q-1)$ and $\mu = \lambda^{-1} \bmod n$.  The
  public key is $\pk = n$ while the secret key is
  $\sk =(n,\lambda,\mu)$.  $c \gets \Enc(\pk, x)$ picks a uniformly
  random integer $\rho \gets [1,n)$ and returns
  $c = (1+x\,n) \rho^n \bmod n^2$.  $x \gets \Dec(\sk, c)$ returns
  $x = L(c^{\lambda} \bmod n) \cdot \mu \bmod n$ where
  $L(a) = (a-1) / n$.  The scheme is additively homomorphic: given
  $c_x \gets \Enc(\pk,x)$ and $c_y \gets \Enc(\pk, y)$,
  $c_x \boxplus c_y = c_x \cdot c_y \cdot \theta^n \bmod n^2$ with
  $\theta \gets [1,n)$.
\end{example}

\paragraph{Oblivious transfer}
A \emph{$1$-out-of-$M$ oblivious transfer (OT) protocol} is a
cryptographic protocol between two parties: a sender and a receiver.
The receiver has an index $j^* \in \{1,\dots,M\}$ as input.  The
sender knows a database $\smallset{x_j}_{j=1,\dots,M}$.  At the end of
the protocol, the receiver learns $x_{j^*}$, while the sender learns
nothing.

As in our protocol $M$ is the number of items in the database, we need
to use practical OT protocols with communication complexity sublinear
in $M$.  We propose to use as $1$-out-of-$M$ OT the basic PIR (Private
Information Retrieval) protocol in~\cite[Sect.~2.2]{PKC:OstSke07}
using the Paillier homomorphic encryption scheme, together with a
classical 1-out-of-$\sqrt{M}$ OT~\cite{SODA:NaoPin01} which is used to
mask the PIR database.  The resulting OT has two rounds (one message
from the receiver to the sender followed by one message from the
sender to the receiver) and its communication complexity is
proportional to $\sqrt{M}$.

\subsection{Matrix factorization}\label{subsec:mfact}
%~~~~~~~~~~~~~~~~~~~~~~~~~~~~~~
The goal of matrix factorization is to predict \emph{unobserved}
ratings $r_{i,j}$ for some user $i$ and some item~$j$, given access to
a set $\TSet$ of user/item pairs $(i,j)$ for which a rating
$r_{i,j} \in \bbbr$ has been generated.  Matrix factorization provides
$d$\nobreakdash-dimensional vectors $\vec{u_i}, \vec{v_j} \in \bbbr^d$
such that
\begin{equation}\label{eq:prediction}
  r_{i,j} \approx \hat{r}_{i,j} := \langle \vec{u_i}, \vec{v_j}\rangle
  = \sum_{k=1}^d u_{i,k} \, v_{j,k}\quad\text{for
    $(i,j) \in \TSet$}\enspace.
\end{equation}
This allows the analyst to predict missing ratings (\ie, those with
$(i,j) \notin \TSet$).  Vector $\vec{u_i}$ is referred to as the
\emph{profile of user~$i$} while vector $\vec{v_j}$ as the
\emph{profile of item~$j$}.

\subsection{Learning the profile of a user}
%~~~~~~~~~~~~~~~~~~~~~~~~~~~~~~~~~~~~~~~~
% These vectors are obtained through the following %regularized
% least-squares estimation,
% \begin{equation*}% \label{eq:MF-LSE}
%   \argmin_{\{\vec{u_i} \in \bbbr^d\}_{1\le i \le N},\\ \{\vec{v_j}
%   \in \bbbr^d\}_{1\le j \le M}} 
%   \sum_{(i,j) \in \TSet} \bigl(r_{i,j} - \langle \vec{u_i}, \vec{v_j} 
%   \rangle\bigr)^2\enspace.
%   % + \lambda\! \sum_{1\le i\le N}{\|\vec{u_i}\|}^2 + \mu\!
%   % \sum_{1\le j\le M}{\|\vec{v_j}\|}^2\enspace. 
% \end{equation*}

% The resulting profiles, $\vec{u_i}$'s and
% $\vec{v_j}$'s, are then used by the analyst to predict missing ratings
% (\ie, those with $(i,j) \notin \TSet$) as
% \begin{equation}\label{eq:prediction}
%   \hat{r}_{i,j} := \langle \vec{u_i}, \vec{v_j}\rangle
%   = \sum_{k=1}^d u_{i,k} \, v_{j,k}\thinspace,
% \end{equation}
% letting $\vec{u_i} = (u_{i,1}, \dotsc, u_{i,d})$ and $\vec{v_j} =
% (v_{j,1}, \dotsc, v_{j,d})$. 

Specifically, when a new user wishes to use the service, she submits a
batch of $s$ ratings $\smallset{r_j}_{j\in \SSet}$ for a subset
$\SSet$ of $s \ge d$ items.  Upon receiving these ratings, the analyst
can estimate her profile $\vec{u}$ through the following least-squares
estimation,\footnote{To ease the presentation, linear regression is
  considered but the proposed techniques readily apply to the more general
  setting of ridge regression.}
\begin{equation}\label{eq:estimate_ui}
  \argmin_{\vec{u}\in\bbbr^d} \sum_{j \in \SSet}\bigl(r_{j} - \langle
  \vec{u}, \vec{v_j} 
  \rangle\bigr)^2\thinspace, 
  % + \nu \, {\|\vec{u_i}\|}^2\thinspace, 
\end{equation}
% for some constant $\nu \ge 0$.
and subsequently predict ratings for items $j \notin \SSet$, using
Eq.~\eqref{eq:prediction}.

Defining matrix $V_\SSet = \left(
  \begin{array}{c|c|c}
    \vec{v_{j_1}} & \hdots & \vec{v_{j_s}}
  \end{array}\right) \in \bbbr^{d \times s}$
and column vector
$\vec{r} = {\left(r_{j_1}, \dots, r_{j_s} \right)}^\intercal \in
\bbbr^s$, the profile $\vec{u}$ of a user can be computed as follows:
\begin{equation}\label{eq:lse-solve}
  \vec{u} = (V_S \cdot V_S^\intercal)^{-1} \cdot V_S \cdot \vec{r} =
  \Biggl(\sum_{k=1}^s \vec{v_{j_k}} \cdot
    \vec{v}_{\vec{j_k}}^\intercal\Biggr)^{-1} \cdot \Biggl(\sum_{k=1}^s
    r_{j_k} \cdot \vec{v_{j_k}} \Biggr)\enspace.  
\end{equation}

%%%%%%%%%%%%%%%%%%%%%%%%%%%%%%%%%%%%%%%%%%%%%%%%%%%%%%%%%%%%%%%%%%%%%%%%%%%%%%%
\section{Our learning protocol}\label{sec:ours}
%%%%%%%%%%%%%%%%%%%%%%%%%%%%%%%%%%%%%%%%%%%%%%%%%%%%%%%%%%%%%%%%%%%%%%%%%%%%%%%

We design a two-round learning protocol between a privacy-conscious
user~$i$ and an analyst, allowing the user to learn her profile
$\vec{u}$ from her (private) ratings $\smallset{r_{j}}_{j \in \SSet}$,
where $\SSet = \smallset{j_1,\dots,j_s}$.  At the end of the protocol,
the analyst will learn nothing (except the size $s$ of $\SSet$), while
the user will only learn her profile $\vec{u}$ and nothing else about
the analyst database (except the dimension $d$, the database of items
and its size $M$, and bounds $B_r$ and $B_v$ on entries of ratings
$r_j$ and of profiles of items $v_j$, respectively).

We insist that our protocol hides the set of actual items $\SSet$ that
the user is rating as they might already leak significant information
about her.  If an upper bound $S$ on $|\SSet|$ is known, the exact
size $s$ of $\SSet$ can trivially be masked by adding fake items (with
profile $\vec{0}$ and fake rating $0$) so that the protocol always
uses a set $\SSet$ of size $S$.

\subsection{Protocol}
%~~~~~~~~~~~~~~~~~~~~
Consider the ring $\Z_n$.  We assume that $n$
is either a prime or is hard to factor, so that for all intents and
purposes $\Z_n$ behaves as a field (since a non-zero non-invertible
element of $\Z_n$ would yield a factor of $n$).  Up to using fixed point
arithmetic (\eg, by multiplying values by some integer $2^\ell$), we
suppose that the entries of $V_S$ and $\vec{r}$ are integers, and so
can be considered as elements of $\Z_n$.

% We remark however that $p$ can be replace by a modulus $n$ (product
% of two large primes) as in the Paillier encryption scheme, as $\Z_n$
% ``behaves'' almost as a field: a non-zero non-invertible element of
% $\Z_n$ allows to factor $n$, which is hard if the Paillier
% encryption scheme is semantically secure.

\paragraph{Round 1}
The user generates a key pair $(\pk,\sk) \gets \KeyGen(1^\secpar)$ for
the homomorphic encryption scheme and encrypts her ratings $r_{j_k}$:
$c_k \gets \HEnc{r_{j_k}}$ for $1 \le k \le s$. She also initiates
$s$~independent OT protocols as a receiver with respective selection
indexes $j_1,\dots,j_s$.

\paragraph{Round 2}
The analyst generates and computes the following matrices (over $\Zn$
and over the ciphertext space respectively):\footnote{We slightly
  abuse notation here.  For vectors, the bracket notation and
  $\boxplus$ and $\boxdot$ operators are applied component-wise.}
\begin{align*}
  A_{k,j}
  &= R_0 \cdot \vec{v_{j}} \cdot \vec{v}_{\vec{j}}^\intercal +  R_k\thinspace,
  &\HEnc{\vec{\alpha_{k,j}}}
  &= (R_0  \cdot \vec{v_{j}}) \boxdot c_{k} \boxplus
    \HEnc{\vec{\rho_k}}\thinspace, 
\end{align*}
where $\smallset{R_k}_{1\le k\le s}$ and
$\smallset{\vec{\rho_k}}_{1 \le k \le s}$ are uniformly random
matrices and vectors summing up to zero in $\Zn^{d \times d}$ and
$\Zn^d$ respectively, and $R_0$ is a uniform matrix in $\GL(d,\Zn)$
(the group of invertible matrices in $\Zn^{d \times d}$).

The analyst then answers the $k$-th OT message from the user, as an OT
sender with database
$\bigl\{x_{k,j} = (A_{k,j}, \HEnc{\vec{\alpha_{k,j}}})\bigr\}_{1 \le
  j\le M}$.

\paragraph{Final step}
The user receives
$x_{k,j_k} = (A_{k,j_k}, \HEnc{\vec{\alpha_{k,j_k}}})$ through the OT
protocols.  We write $A_k = A_{k,j_k}$ and
$\vec{\alpha_k} = \Dec(\sk, \HEnc{\vec{\alpha_{k,j_k}}})$.  We then
remark that, in $\Zn$:
\[
  (V_S \cdot V_S^\intercal)^{-1} \cdot V_S \cdot \vec{r} =
  \Biggl(\sum_{k=1}^s A_k \Biggr)^{-1} \cdot \Biggl(\sum_{k=1}^s
  \vec{\alpha_k} \Biggr)\enspace.
\]
So if $n$ is large enough, the user can compute back $\vec{u}$ using
rational reconstruction~\cite{WanGuyDav82,FC:FouSteWac02} (we recall
that $\vec{u}$ satisfies \cref{eq:lse-solve} over the rationals, and
that $\vec{u}$ is not necessarily an integer).

\paragraph{Bounds for correctness}
The scheme is correct when the above rational reconstruction succeeds.
From~\cite{WanGuyDav82,FC:FouSteWac02} and Hadamard's
inequality, we can show correctness when
$n > 2d^{d+1/2} s^{2d+1} B_V^{4d+1} B_r$, where $B_V$ and $B_r$ are
upper bounds on the absolute values of the coefficients of the item
profiles $\vec{v_j}$ and of the ratings $r_j$, respectively.
For example, if $B_V = 2^{20}$, $B_r = 4$, $d=8$, $s = 10$, this is
already satisfied for an integer $n$ of $806$ bits.

\paragraph{Security}
Security against semi-honest adversaries follows from the security of the OT protocol, the IND-CPA property of the homomorphic encryption scheme, and from the following fact:
since $V_S \cdot V_S^\intercal$ is invertible and
$\GL(d,\Zn)$ is a group, %given access to $\vec{u}$ only,
% it is
%possible to generate $\smallset{A_k}_k$, $R$,
%$\smallset{\vec{\alpha_k}}_k$, and $\vec{\rho}$, according to the same
%distribution as above.  That means that
$\smallset{A_k}_k$ and $\smallset{\vec{\alpha_k}}_k$ only reveal $\vec{u}$.

% \subsection{Setting}
% %~~~~~~~~~~~~~~~~~~~
% \todo[inline]{Recall that giving the ratings in the clear is leaking
%   too much info.  For this reason, we adopt the setting
%   of~\cite{CCS:NIWJTB13}.  Also detail how to compute rating in a
%   privacy-preserving fashion; cf. Section~4.1 in~\cite{CCS:NIWJTB13}.}

\subsection{Instantiation using Paillier homomorphic encryption scheme}
%~~~~~~~~~~~~~~~~~~~~~~~~~~~~~~~~~~~~~~~~~~~~~~~~~~~~~~~~~~~~~~~~~~~~~
The scheme can be instantiated using the Paillier encryption scheme
and the OT described in \cref{sec:preliminaries}.  We can use the
internal construction of the OT, to avoid sending ciphertexts of
$r_{j_k}$.  Concretely, in the OT construction, the user encrypts a
vector $(0,\dots,1,\dots,0)$ used to ``select'' the correct value to
be received.  If we use two OT protocols for each $k$, one for $A_k$
and one for $\vec{\alpha_{k}}$ (instead of a single one for the pair
$(A_k,\vec{\alpha_k})$), then for the second OT, the user just
encrypts $r_{j_k}$ instead of $1$, she will receive $r_{j_k}$ times
the value to be received.

The resulting protocol for $M = 100$ items, dimension $d=8$, and
$s = 10$ ratings from the user (modulus $n$ of size $1024$ bits for
Paillier encryption scheme and an elliptic curve over a 256-bit prime
field for the base OT~\cite{SODA:NaoPin01}), has the following
performance on a non-optimized single-thread implementation (on a
laptop, CPU Intel\textregistered{} i7-7567U, $3.5$GHz, turbo $4$GHz):
less than $0.4$s to generate the first round by the user, less than
$150$s to generate the second round by the analyst, less than $1.4$s
to finalize the protocol by the user. The user requires less than $2$s
of computation (excluding communication). The analyst time is mostly
spent in the exponentiations required in the OT protocol (modulo
$n^2$): there are $M \cdot (d^2 + d) \cdot s$ of them. These
exponentiations can be trivially parallelized.  The communication
complexity is less than $2$MB and essentially grows linearly with
$\sqrt{M}$.

% \bibliographystyle{plain}
% \bibliography{cryptobib/abbrev3,add,cryptobib/crypto}

\providecommand{\noopsort}[1]{}

\end{document}